\pdfoutput=1
\documentclass[a4paper,11pt]{article}
\usepackage{pos}
\usepackage{wrapfig}
\usepackage{graphicx}
\usepackage{subcaption}
\usepackage{xcolor}
\usepackage{aas_macros}

\title{GeV–TeV Connections in Galaxies: Evolutionary Signatures from Pulsars in Globular Clusters}
\ShortTitle{GeV–TeV Connections in Galaxies}

\author*[a]{Ellis R. Owen}
\author[b,c,d]{Yoshiyuki Inoue}
\author[e]{Chung-Yue Hui}
\author[b]{Tatsuki Fujiwara}
\author[f]{Albert K. H. Kong}

\affiliation[a]{Astrophysical Big Bang Laboratory (ABBL), RIKEN Pioneering Research Institute (PRI), Wak\={o}, Saitama, 351-0198 Japan}

\affiliation[b]{Department of Earth and Space Science, Graduate School of Science, The University of Osaka, Toyonaka, Osaka 560-0043, Japan}

\affiliation[c]{RIKEN Center for Interdisciplinary Theoretical and Mathematical Sciences (iTHEMS), Wak\={o}, Saitama 351-0198, Japan}

\affiliation[d]{Kavli Institute for the Physics and Mathematics of the Universe (WPI), UTIAS, The University of Tokyo, Kashiwa, Chiba 277-8583, Japan}

\affiliation[e]{Department of Astronomy \& Space Science, Chungnam National University, Daejeon 34134, Korea}

\affiliation[f]{Institute of Astronomy, National Tsing Hua University, Hsinchu 30013, Taiwan}

\emailAdd{ellis.owen@riken.jp}

\abstract{The dominant mechanisms underlying high-energy $\gamma$-ray emission from galaxies vary by galaxy type. In starbursts, a major contribution comes from neutral pion decay. This is driven by interactions between interstellar gas and hadronic cosmic rays (CRs), which are accelerated in strong shocks associated star formation activity and stellar remnants. Leptonic $\gamma$-ray emission can also arise from electrons directly energized in interstellar shocks, produced via charged pion decays, or emitted by pulsars and their surrounding halos. In quiescent galaxies, pulsars and their halos can represent a major $\gamma$-ray source class, with millisecond pulsars predominantly located in globular clusters (GCs) being particularly important. Recent detections of very high-energy (VHE) emission from Galactic GCs suggests they may also contribute to the TeV $\gamma$-ray flux from evolved galaxies. We consider a scenario where this VHE emission from GCs is powered by electrons accelerated in communal stellar/pulsar wind cluster termination shocks. These electrons undergo inverse Compton scattering as they propagate into GC magnetotails. Our results show that the high-energy emission from GCs can be an important contributor to the GeV and TeV flux from massive, quiescent galaxies. The relative strength of each component depends on the global galactic properties and its evolutionary history.}

\ConferenceLogo{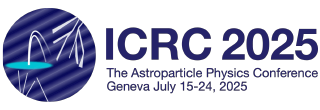}

\FullConference{39th International Cosmic Ray Conference (ICRC2025)\\
 15–24 July 2025\\
Geneva, Switzerland\\}

\begin{document}
\maketitle

\section{Introduction}
\label{sec:intro}

\vspace{-0.3cm}
\noindent 
Globular clusters (GCs) in our Galaxy are established GeV $\gamma$-ray sources. To date, associations have been reported for 45 GCs in \textit{Fermi}-LAT data~\citep{2023arXiv230712546B, Shin2025A&A}. Their high-energy emission is typically attributed to cosmic ray (CR) electrons accelerated in the magnetospheres of millisecond pulsars (MSPs) residing in these clusters. This interpretation is supported by the similarity between the $\gamma$-ray spectra of most GCs and that of individual MSPs
~\citep[e.g.][]{GC_LAT}, as well by the detection of millisecond $\gamma$-ray pulsations from some clusters~\citep[e.g.][]{Wu2013ApJ}. 

MSPs form via a spin-up process. In the Galactic field, this occurs through accretion in compact binaries~\citep{Alpar1982Natur}. In dense stellar environments like GC cores, correlations observed between GC $\gamma$-ray luminosity (or MSP number) and the stellar encounter rate~\citep[e.g.][]{GC_LAT} indicate that dynamical encounters provide an additional spin-up channel. 
MSPs have low spin-down rates~\citep{Hui2019Galax}, allowing them to accelerate CR electrons steadily over Gyr timescales. As a result, their $\gamma$-ray output is long-lived and persistent, continuing long after their formation. 

This sustained emission from MSP populations may make an important contribution to the high-energy output of evolved galaxies. 
For instance, MSPs have been proposed as a possible origin of the GeV excess observed toward the Galactic Center and M31~\cite[e.g.][]{Eckner2018ApJ}. Indeed, M31 hosts more than 460 GCs~\cite[e.g.][]{Barmby2001AJ}. If each cluster has a GeV-band luminosity comparable to Terzan 5~\cite{Song_2021}, their combined $\gamma$-ray luminosity could reach $\sim 2\times 10^{38}$ erg s$^{-1}$. 
This is close to the theoretical maximum energy budget available in M31 for $\gamma$-ray emission from hadronic pion decays,\footnote{Assuming a CR calorimetric fraction of 5\% \cite{Roy2022MNRAS} and negligible CR cooling losses.} $\sim 2\times 10^{39}$ erg s$^{-1}$, demonstrating that
GCs can be a competitive galactic $\gamma$-ray source. 

CR electrons injected by MSPs in GCs may undergo further acceleration at bow shocks. These shocks develop as MSP and stellar winds interact with the surrounding ISM as the cluster moves through its host galaxy~\cite{Bednarek2007MNRAS}. The re-accelerated electrons can up-scatter ambient photons, typically from the cosmic microwave background or the local ISRF, via inverse Compton scattering (ICS)~\citep{Bednarek2007MNRAS}. This process produces very high energy (VHE) TeV $\gamma$-rays, which 
have been confirmed observationally in Terzan 5 by H.E.S.S.~\cite{HESS2011A&A}, with a further possible association recently reported in the direction of UKS 1~\citep{Shin2025A&A}. 

The GeV-TeV emission spectrum from a GC is governed by several environmental parameters: the GC’s MSP and stellar luminosity~\citep{Shin2025A&A}, the strength of the local ISRF~\citep[e.g.][]{Cheng_2010}, and the ambient ISM density~\citep[][]{Bednarek_2014}. Since these factors vary with galactocentric radius, the VHE $\gamma$-ray output of a galaxy’s GC system is expected to depend on the spatial distribution of its clusters. This distribution is generally governed by a galaxy’s assembly history. 
Thus, the relationship between a galaxy’s integrated GeV and TeV $\gamma$-ray emission and its GC population may provide information about its past evolution. In this study, we investigate how different galaxy formation scenarios imprint distinct spectral signatures on the $\gamma$-ray emission from GC systems, and examine their potential as tracers of galactic evolutionary pathways.

\vspace{-0.3cm}
\section{GC systems in evolving galaxies}
\label{sec:gc_systems}

\noindent 
GC systems in galaxies often show a bimodal metallicity distribution. These are metal-poor (MP or ``blue'') clusters and metal-rich (MR or ``red'') clusters~\cite{Brodie2006ARA&A}. MP clusters are typically found at larger galacto-centric radii, residing in the galactic halo, while MR clusters tend to be more centrally concentrated. 
The formation pathways of these two populations are generally different. MP clusters form in low-mass halos that produce satellite galaxies. These later accrete onto more massive galaxies in minor mergers, favouring the deposition of MP clusters in galaxy outskirts. MR clusters are typically formed during gas-rich (or ``wet'') major mergers or strong galaxy tidal interactions. In these events, rapid gas accretion triggers intense GC formation \textit{in situ}, especially in the central regions of the resulting post-merger galaxy. 

\vspace{-0.2cm}
\subsection{MSP formation in MP and MR clusters}
\label{sec:msps_in_gcs}

\noindent 
MSPs are recycled pulsars. In dense stellar environments such as GCs, they mainly form by dynamical interactions, particularly via neutron star capture during close stellar encounters. 
The encounter rate can be written as: 
\begin{equation}
\Gamma_{\rm c} \simeq 13.7 \;\! \left( \frac{\rho^{\star}}{10^4 \; {\rm L}_{\odot} \;\! {\rm pc}^{-3}} \right)^2 \;\! \left( \frac{r_{\rm c}}{1.0 \;\! {\rm pc}} \right)^3 \;\! \left( \frac{\sigma_0}{10.0 \;\! {\rm km} \;\! {\rm s}^{-1}} \right)^{-1} \ , 
\label{eq:dynamical_formation_msps}
\end{equation}
(scaled from Ref.~\cite{Hui2010.714}). Here, $\rho^{\star}$ is the central luminosity density of the cluster, $r_{\rm c}$ is the core radius, and $\sigma_0$ is the cluster's stellar velocity dispersion. 
If the encounter rate remains approximately constant over a cluster's lifetime, the number of MSPs is expected to scale with it.\footnote{Observations have confirmed this correlation, however additional dependencies on cluster metallicity have also been found~\cite[e.g.][]{Hui_2011}.  
This may arise because magnetic braking is more effective in metal-rich clusters, which boosts orbital decay in binaries. This increases the likelihood of MSP formation~\cite{Hui_2011}. More recent studies have suggested that correlations with metalicity may be weaker \cite{Menezes2019MNRAS}.}  
The number of MSPs $N_{\rm MSP}$ in each GC can be estimated using  $\log N_{\rm MSP} = 0.6 \;\! [{\rm Fe}/{\rm H}] + 0.39 \log \Gamma_{\rm c} + 1.78$. 
We apply this relation separately to MP and MR GC populations, adopting representative metallicities of [Fe/H] = –1.5 for MP clusters and [Fe/H] = –0.5 for MR clusters~\cite{Brodie2006ARA&A}.  The resulting GeV $\gamma$-ray luminosity is then $L_{\gamma} = N_{\rm MSP} \langle \dot{E} \rangle \langle \eta_{\gamma} \rangle$, where we adopt $\langle \dot{E} \rangle = 2\times 10^{34} \;\! {\rm erg}\;\! {\rm s}^{-1}$ for the average MSP spin-down power, and
$\langle \eta_{\gamma} \rangle \sim 0.08$ for the average $\gamma$-ray production efficiency~\cite{Abdo2010A&A}. We consider that the corresponding TeV emission for a GC is driven by ICS, following~\cite{Bednarek2007MNRAS}. 

\vspace{-0.2cm}
\subsection{GC distributions}
\label{sec:gc_distribution_model}

\noindent 
The radial distribution of GC systems in galaxies typically follows a power-law surface density profile. 
In the outer halo, the profile declines steeply~\cite{Brodie2006ARA&A}. 
To capture this, we construct a simple parametric model for the distribution of GCs in a galaxy, where the total GC number density is given by: 
\begin{equation}
n_{\mathrm{GC}}(r, t) = \sum_{i = \mathrm{MP,\,MR}} N_i(t)\,
\left( \frac{r}{r_i(t)} \right)^{-\alpha_i(t)} 
\exp\left[ -\left( \frac{r}{r_{\mathrm{cut},i}} \right)^{\beta_i} \right] \ ,
\label{eq:gc_syst_evo}
\end{equation}
\noindent
and where the two components $i$ correspond to the MP (halo-tracing) and MR (bulge-tracing) populations. Here, $\alpha_i$ is the inner slope of the density profile for each component, $\beta_i$ sets the exponential suppression at large radii above a $r_{\mathrm{cut},i}$ truncation radius, $r_i(t)$ is the scale radius of component $i$, which may evolve over time, and $N_i(t)$ is the total number of GCs in component $i$ at time $t$. This evolves according to the host galaxy's merger history (see section~\ref{sec:merger_histories}). 

\subsection{Other MSP populations in galaxies}
\label{sec:msp_general}

\noindent
MSPs can also form via dynamical interactions in galactic bulges, where stellar densities may reach levels comparable to those in some GCs~\cite{Eckner2018ApJ}. We model their formation using the same prescription applied to GC MSPs (Eq.~\ref{eq:dynamical_formation_msps}, with $r_{\rm c}$ set as the bulge scale radius $r_{\rm b}$). MSPs also form in the galactic field through accretion of material onto neutron stars in close binary systems. Their population development is modeled using the approach of Ref. \cite{Owen2025arXiv}. We include the GeV-band contribution of both bulge and field MSPs. 
These MSPs could also produce TeV-band ICS emission~\cite[e.g.][]{Bednarek2013A&A}, but we find this contribution is sub-dominant. 

\vspace{-0.3cm}
\section{$\gamma$-ray emission from galaxies under different evolutionary pathways}
\label{sec:merger_histories}

\noindent
We adopt a prototype-based approach to model the high-energy emission from galaxies with different evolutionary histories. In all cases, we use parametric profiles to describe both the gas density distribution and ISRF, which are main inputs that govern the TeV emission from a galaxy's GC system. Each profile consists of a compact bulge and an extended halo component: 
\begin{equation}
    n_{\mathrm{gas}}(r) = 
    n_0^{\mathrm{b}}\, \exp\left(-\frac{r}{r_{\mathrm{b}}}\right)
    + n_0^{\mathrm{h}} \left( 1 + \frac{r^2}{r_{\mathrm{h}}^2} \right)^{-3\beta/2} \,,
\end{equation}
where $n^{\rm b}_{\rm 0}$ and $n^{\rm h}_{\rm 0}$ are the central gas densities of the bulge and halo, respectively. The parameters 
$r_{\rm b}$ and 
$r_{\rm h}$ are the bulge scale radius and halo core radius, while
$\beta$ sets the slope of the halo profile. The ISRF is modeled in an analogous form, substituting gas densities with radiation energy densities ($u^{\rm b}_{\rm 0}$ and $u^{\rm h}_{\rm 0}$ for the bulge and halo, respectively).\footnote{Post-merger galaxies are expected to have increased ISRF energy densities, however the precise ISRF spectrum does not greatly affect our results and is not explicitly modeled in this work.}

Following a major merger, galaxies are expected to experience compaction. This produces smaller, denser bulges and more extended halos with shallower gradients. Minor mergers typically produce less severe structural changes, while quiescent systems generally retain diffuse halos and low-density bulges with steeper radial profiles. 

\subsection{Evolving galaxy scenarios}
\label{sec:galaxy_scenarios}

\noindent
\textbf{Quiescent:} In quiescent galaxies without recent significant merger events, GCs are gradually disrupted over time due to relaxation, tidal stripping, disk/bulge shocking, and dynamical friction. We model these effects using a characteristic GC dissipation rate:
\begin{equation}
    \frac{{\rm d}N_i(t)}{{\rm d}t} = - \frac{N_i}{\tau_{{\rm dis}}(r)} \ , 
\end{equation}
where both MP and MR GC populations dissolve over the same timescale. The cluster dissipation timescale, $\tau_{\rm dis}$, is modeled as a function of GC mass and local ambient density~\cite{Lamers2005A&Ab}: 
\begin{equation}
\tau_{\rm dis} \approx 810 \left( \frac{M_{\rm GC}}{10^4~{\rm M}{\odot}} \right)^{0.62} \left(\frac{\rho}{1 \;\! \rm{M}_{\odot} {\rm pc}^{-3}}\right)^{-1/2} ~{\rm Myr} \ , 
\end{equation}
where $M_{\rm GC}$ is the GC mass (we adopt a mean mass of $10^4 \;\!{\rm M}_{\odot}$, with more detailed treatments left to follow-up work) and $\rho$ as the density of the ambient ISM gas. 
 The power-law slope of the GC radial distribution does not evolve in the quiescent case, and is fixed at $\alpha_i = -3.5$ for both MP and MR populations~\cite[cf. pre-merger systems in Ref.][]{Bekki2006A&A}.

\vspace{0.2cm}
\noindent 
\textbf{Strong Interaction / Major merger:} Major mergers, particularly gas-rich (``wet'') mergers, and close galaxy-galaxy interactions can trigger intense starburst activity, leading to \textit{in-situ} formation of MR clusters at a rate: 
\begin{equation}
    \frac{{\rm d}N_{\mathrm{MR}}(t)}{{\rm d}t} = 
    \epsilon_{\mathrm{GC}}\, \dot{R}_{\mathrm{int}} \, \frac{M_{\mathrm{gas}}}{M_{\rm GC}} \,
\end{equation} 
where 
$\epsilon_{\rm GC} (= 0.2 \%)$ is the GC formation efficiency,  
$\dot{R}_{\rm int}$ is the galaxy-galaxy interaction or major merger rate, and $M_{\rm gas}$ is the mass of gas involved in the starburst. The MP population is unaffected. 

Newly formed MR clusters are initially centrally concentrated, following the stellar light profile of the post-merger remnant. However, violent relaxation and dynamical mixing associated with these disruptive events act to flatten the distribution. The radial index of the GC profile is therefore modeled to increase by 0.3 per merger~\cite{Bekki2006A&A}, from an initial value of $-3.5$. We consider that the overall scale and truncation radius of the GC system remain unchanged.

\vspace{0.2cm}
\noindent 
\textbf{Minor merger:} In galaxies undergoing minor mergers, the GC system evolves through internal dissipation  
and the accretion of GCs from in-falling satellite galaxies~\cite{Brodie2006ARA&A}. These accreted GCs typically formed at early epochs and are predominantly metal-poor. We model the supply of MP clusters via minor mergers as: 
\begin{equation}
    \frac{{\rm d}N_{\mathrm{MP}}(t)}{{\rm d}t} = 
    f_{\mathrm{sat}}\, \dot{R}_{\mathrm{min}}\, N_{\mathrm{GC,sat}} \,,
\end{equation}
with no contribution to the MR population. Here, $f_{\rm sat}$ is the fraction of satellite GCs that survive and are retained by the host after a minor merger,\footnote{The value of $f_{\rm sat}$ is not well constrained; we set its value such that the final fraction of GCs formed via satellite accretion matches results from numerical simulations~\cite{Keller2020MNRAS}.} $\dot{R}_{\rm min}$ is the minor merger rate, and $N_{\rm GC, sat}$ is the typical number of GCs hosted by a satellite galaxy.

Satellite galaxies typically merge into the outskirts of their host. Therefore MP clusters are preferentially deposited at large radii. This leads to a broadening of the GC system, which we model as:
\begin{equation}
    \frac{{\rm d}r_{\mathrm{MP}}(t)}{{\rm d}t} = 
    f_{\mathrm{sat}}\, \Delta r\, \mu\, \dot{R}_{\mathrm{min}} \,,
\end{equation} 
where $r_{\rm MP}(t)$ is the scale radius of the GC system (see eq.~\ref{eq:gc_syst_evo}), $\Delta r$ is the 
typical offset of in-falling satellites, and $\mu$ is the average mass ratio of minor mergers. 
The preferential supply of GCs at large radii in minor mergers also flattens the radial profile~\cite{Bekki2006A&A}. We account for this by time-evolving the power-law slope at 10\% of the rate of major mergers.   

\vspace{-0.2cm}
\subsection{Emission from hadronic and leptonic CRs}
\label{sec:had_lep_emission}

\noindent 
In addition to $\gamma$-ray emission from GC systems, high-energy emission can also originate from CRs energized within galaxies by shocks and turbulence. These are usually associated with star-formation activities.  Above a threshold energy of $\sim$280 MeV, hadronic CRs interact with ambient gas to produce charged and neutral pions. These decay, producing $\gamma$-rays, neutrinos and secondary electrons/positrons. Additional $\gamma$-ray emission arises from bremsstrahlung and ICS of primary CR electrons, as well as the secondaries produced by charged pion decays. These contributions are modeled using the approach in Ref.~\cite{Owen2025arXiv}.

\section{Results, discussion and conclusions}
\label{sec:results_discussion}

\begin{figure}
    \centering
    \includegraphics[width=\linewidth]{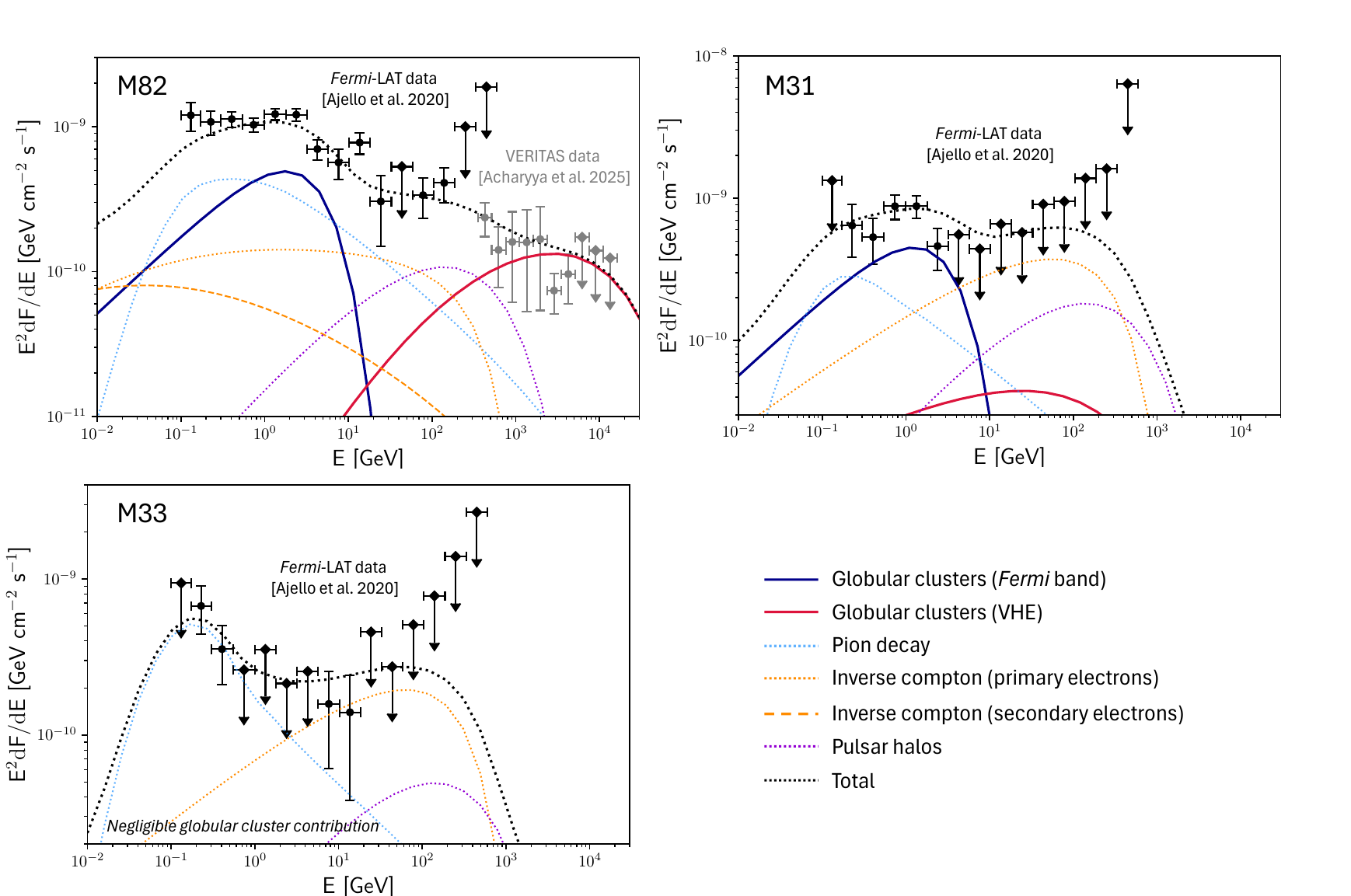}
    \caption{Model $\gamma$-ray spectra for typical parameters adopted for each galaxy (without fitting), compared with \textit{Fermi}-LAT data from Ajello et al. 2020 (Ref.~\cite{Ajello2020ApJ}).  \textbf{Strong interactions: M82} is undergoing a sustained interaction with M81. We treat this as an ongoing interaction activity, equivalent to a major event every Gyr. \textbf{Minor mergers: M31} is modeled with 3 minor mergers per Gyr. \textbf{Secular growth: M33} has experienced no major mergers in the last few Gyrs, with no significant minor merger activity. While some GCs do form, their contribution to the overall $\gamma$-ray emission is negligible.}
    \vspace{-0.5cm}
    \label{fig:fig1}
\end{figure}

\vspace{-0.3cm}
\noindent
We consider three nearby GeV $\gamma$-ray-detected galaxies with distinct formation histories: M82, M31 and M33. While their precise merger histories are uncertain, simulations of typical evolutionary pathways for galaxies of their type~\cite[e.g.][]{Sotillo2022MNRAS} provide a useful statistical baseline to inform source-specific constraints. Using representative rates for minor and major mergers/interactions, we model the evolution of their GC populations and predict their GeV-TeV emission spectra. Our results are shown in Fig.~\ref{fig:fig1} in comparison to \textit{Fermi}-LAT spectral data from Ref.~\cite{Ajello2020ApJ}.

Our results show that GeV and TeV $\gamma$-ray emission can serve as a tracer of a galaxy's formation history, particularly in systems with low current star formation. Galaxies that experienced a recent major merger or strong interaction are expected to present strong GeV and TeV emission features dominated by GCs in their bulge, where higher metallicity favours MSP formation. Galaxies that evolved through frequent minor mergers (without major mergers or interactions) can still develop a rich GC population, particularly in their outskirts. However, low gas densities and weak, distant bow shocks in these outer regions limit the efficiency of TeV ICS emission. These galaxies are therefore likely to display strong GeV emission from MSPs in GCs but lack any strong TeV component. Quiescent galaxies built primarily through gradual gas accretion are unlikely to produce significant GC-related high-energy emission. In these cases, the observed $\gamma$-ray flux is likely dominated by other processes, e.g. hadronic/leptonic CR emission associated with star formation or AGN activity.

This work is a first parametric study of the high-energy emission from GC systems in galaxies. Although it does not include detailed spectral modeling or advanced treatments of particle acceleration, it demonstrates that GCs can make a substantial contribution to a galaxy’s GeV–TeV emission. This emission can persist for several Gyr after the formation of the GC population, and encodes valuable information about the host galaxy’s merger history. This offers a new window to trace galaxy evolution that can complement conventional approaches (e.g. reaching into dense regions like the bulge that are often obscured). 
As the number of GeV-detected galaxies increases, identifying GC-related signatures will become increasingly viable. Targeted TeV follow-up observations will offer a powerful means to distinguish between different evolutionary pathways and constrain the assembly histories of nearby galaxies. 

\vspace{0.1cm}
\noindent
\footnotesize
\textbf{Acknowledgements:} E.R.O. is supported by the RIKEN Special Postdoctoral Researcher Program for junior scientists. Y.I. is supported by an NAOJ ALMA Scientific Research Grant Number 2021-17A, JSPS KAKENHI Grant Numbers JP18H05458, JP19K14772, and JP22K18277, and the World Premier International Research Center Initiative (WPI), MEXT, Japan. C.Y.H. is supported by the research fund of Chungnam National University and by the National Research Foundation of Korea grant 2022R1F1A1073952.
\normalsize

\vspace{-0.4cm}
\bibliographystyle{ICRC}
\bibliography{references}

\end{document}